\documentclass{PoS}
\usepackage{subfig}

\title{ Galactic Cosmic Ray Sun Shadow during the declining phase of
  cycle 24 observed by HAWC}

\ShortTitle{HAWC Sun Shadow}

\author{\speaker{Alejandro Lara},$^{ab}$
  Paulina Colin,$^{ac}$ K. P. Arunbabu $^{a}$ and
  James Ryan$^{d}$ 
  for the HAWC collaboration\footnote{for collaboration list see
    PoS(ICRC2019)1177 or visit https://www.hawc-observatory.org} \\
  \llap{$ˆa$} Instituto de Geof\'isica, UNAM, M\'exico\\
  \llap{$ˆb$} The Catholic University of America, USA\\
  \llap{$ˆc$} Posgrado en Ciencias de la Tierra, M\'exico\\
  \llap{$ˆd$} University of New Hampshire, USA.\\
        E-mail: \email{alara@igeofisica.unam.mx}}

      \abstract{
The High Altitude Water Cherenkov (HAWC) array is sensitive to high energy Cosmic Rays
(CR) in the $\sim 10$ to $\sim 200$ TeV energy range, making it
possible to construct maps of the so called
“Sun Shadow” ($SS$), i. e. of the deficit of CR coming from the direction of the Sun. In this
work, we present the variation of the Relative Intensity of the deficit ($SS_{RI}$)  for three years of
HAWC observations form 2016 to 2018 in which we found a clear decreasing trend of the ($SS_{RI}$)
over the studied period, corresponding to the declining phase of the solar cycle 24. By comparing
the $SS_{RI}$ with the photospheric magnetic field evolution, we show that there is a linear
relationship between the $SS_{RI}$ and the median photospheric magnetic field of the Active Region
belt (-40$^\circ \le$  lat $\le$ 40$^\circ$)  and a inverse linear relationship with the polar photospheric magnetic
field  (lat $\ge \pm$ 60$^\circ$).
 The former relationship is due to the magnetic field causing a deviation of
the CR, whereas the latter reflects the change of the heliospheric field topology from multipolar to
dipolar configurations. These relationships are valid only when the median magnetic field is
lower than 8 G, during the declining and minimum phases of the solar cycle 24. Finally, we show
that relativistic charged particles, in the 10 to 200 TeV energy
range, are deflected a 
few degrees.
}

\FullConference{36th International Cosmic Ray Conference -ICRC2019-\\
		July 24th - August 1st, 2019\\
		Madison, WI, U.S.A.}

\begin{document}

\section{Introduction}

HAWC is an air shower array located at 4100 m above the sea level on the Sierra Negra
Volcano in the central part of Mexico (N 18$^\circ$ 59’ 48”, W 97$^\circ$ 18’ 34”), intended for the exploration
of the Northern hemisphere sky in high energy gamma rays. Although, as the majority of primary
particles (> 99\%) are hadrons, HAWC is a sensitive Galactic Cosmic Rays (GCR) telescope with
an energy range from $\sim 10$ to
$\sim 200$ TeV. A description of HAWC is presented in \cite{2014PhRvD..90l2002A, 2017ApJ...843...39A}.

GCRs are charged particles, mainly protons, with energies ranging form
few  $10^{6}$ up to $10^{20}$   eV
arriving isotropically from outside the heliosphere. The heliospheric magnetic field modulates the
GCR flux that reaches the Earth, making the study of GCR an excellent tool to remotely explore the
Heliosphere. For example, low energy GCRs  (E $< 10^{6}$)  provide information on the outer
Heliospheric Magnetic Field  \cite{1998JGR...103.2099H}  and long term solar modulation \cite{2010JGRA..11512109M}. At higher energies (tens of
GeV) GCRs are well suited for short term solar transient studies, such as coronal mass ejections,
solar flares and high speed steamers  \cite{2009AdSpR..43..480B}. Only high energy GCRs (E $> 10^{12}$ eV) are unaffected by
the interplanetary magnetic fields and therefore, are able to reach
the low atmosphere of the Sun, interacting there with both: the strong
near-photospheric magnetic fields and  particles \cite{1992NASCP3137..542S}  giving in this way, valuable information of these fields. At
the Earth, an observer with a sufficiently sensitive GCR telescope pointing towards the Sun, will
see the deficit of the high energy GCR flux, because of the Sun's physical presence and the
deviation of the GCRs caused by the solar magnetic field.

In this work we use the High Altitude Water Cherenkov Array data to construct maps of the
high energy GCR deficit observed at the solar position, the so called Sun Shadow  ($SS$) maps  (Sec. \ref{sec:ssmaps}).
 Because HAWC has been observing the high energy sky since 2016, we have enough data to
explore the SS with long integration times (giving signal to noise ratios greater than 40) and with
medium integration times corresponding to one Carrington Rotation ($\sim 27$  days). The period of
our study corresponds to the declining phase of Cycle 24 and in Sec. \ref{sec:cycle}. We explore the
photospheric magnetic field evolution during this phase. The physical connection between high
energy GCR and the photospheric magnetic field is explored through a basic simulation in Sec.  \ref{sec:sim}.
Finally, our conclusions are in Sec. \ref{sec:concl}.

\section{Sun Shadow Maps} \label{sec:ssmaps}

\begin{figure}%
    \centering
    \subfloat[2016]{{\includegraphics[width=7cm]{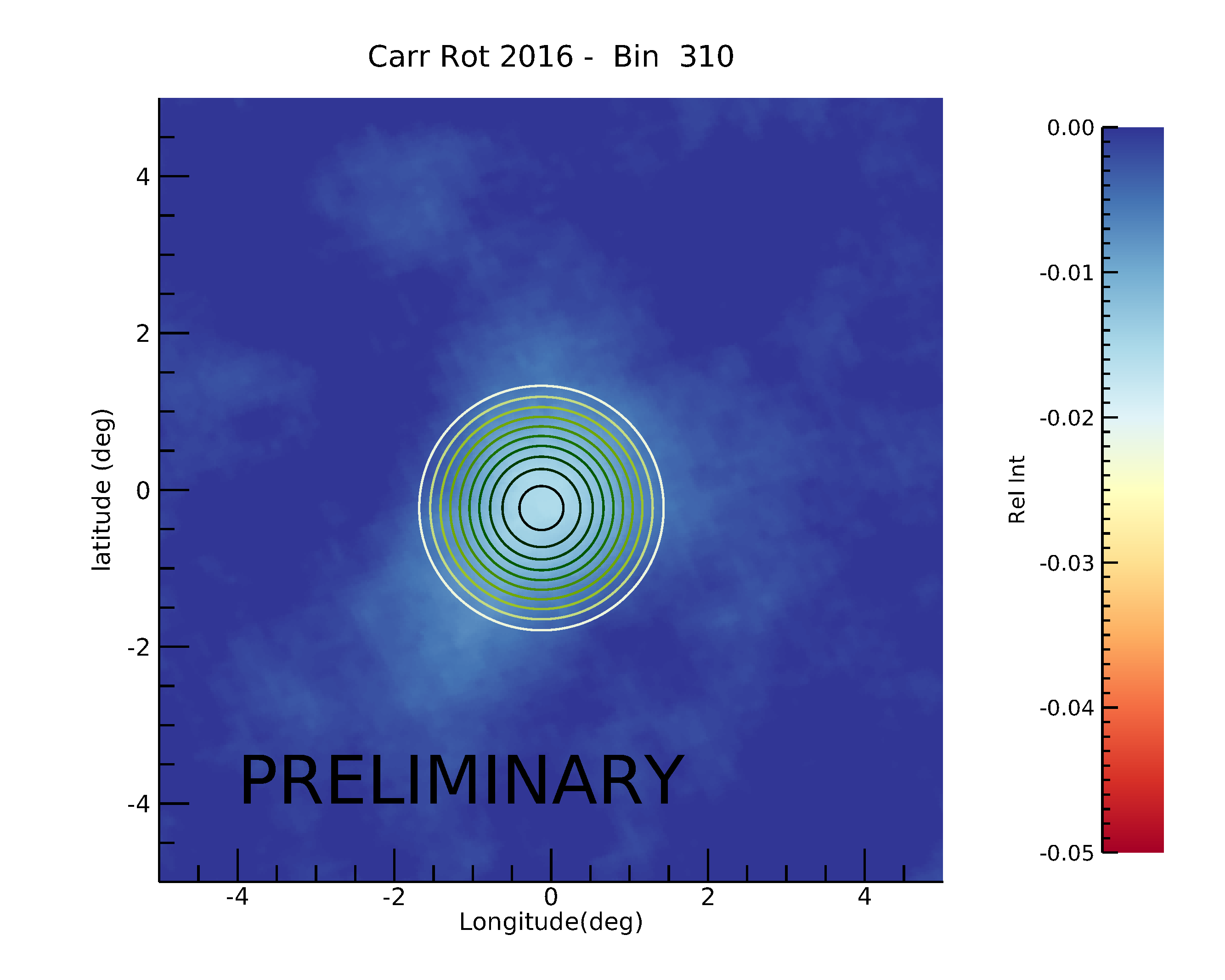} }}%
    \qquad
    \subfloat[2017]{{\includegraphics[width=7cm]{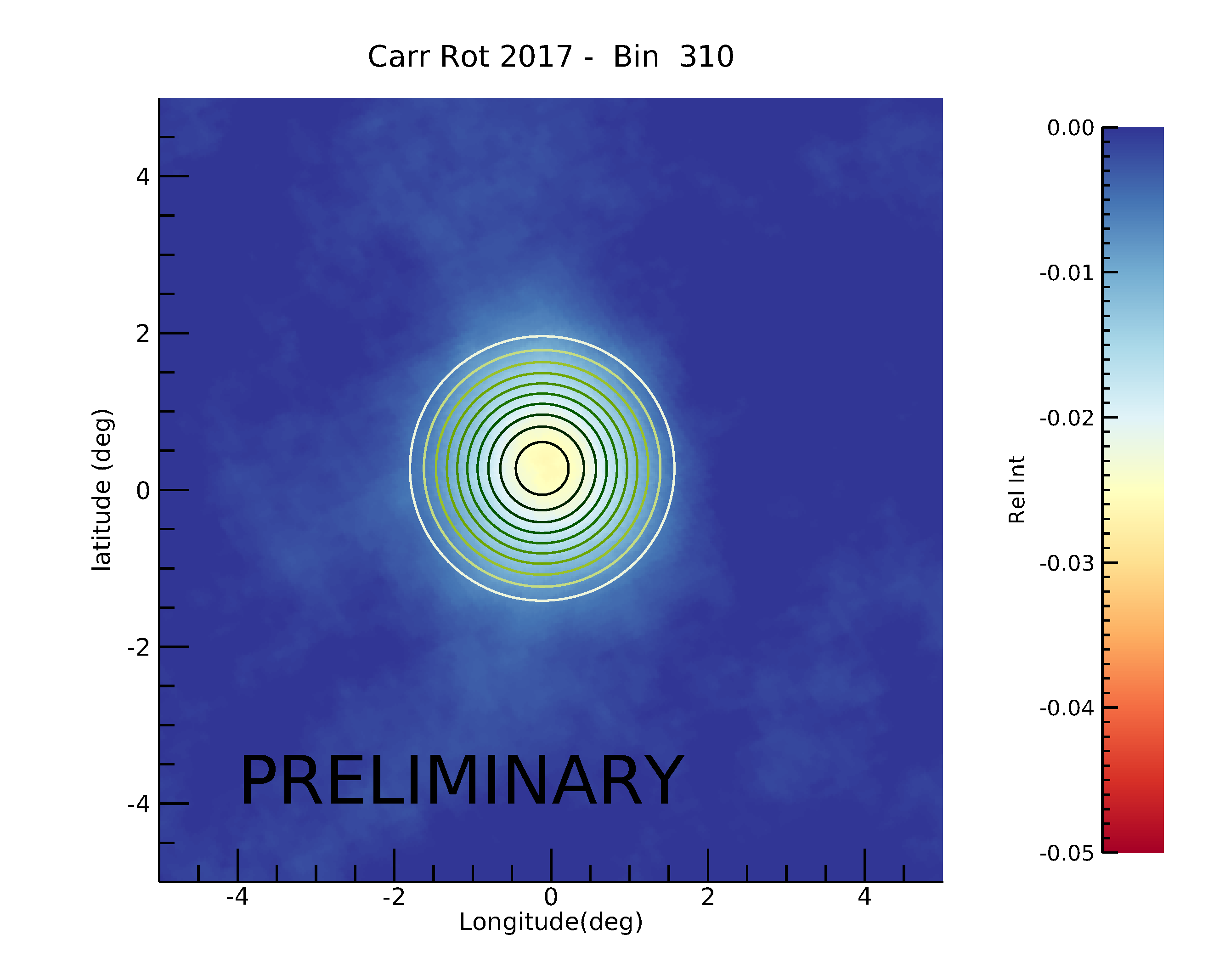} }}%
    \caption{Maps of the relative intensity of the  $SS_{RI}$  integrated
      during 2016 (a) and 2017 (b) The contours correspond to the 2D
      fitted to each map.}%
    \label{fig:maps16y17}%
\end{figure}

\begin{figure}%
    \centering
    \subfloat[]{\includegraphics[width=7cm]{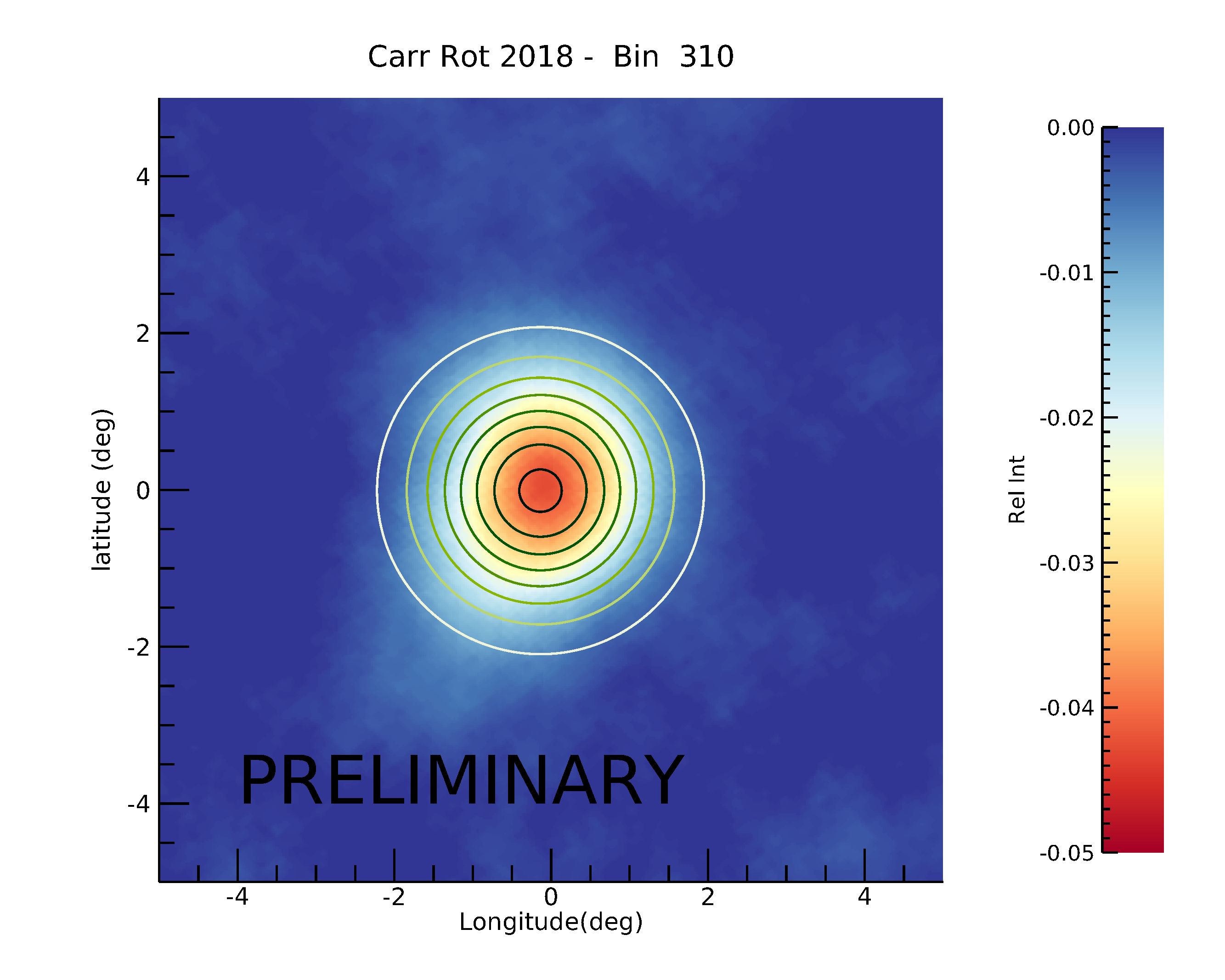}}%
    \qquad
    \subfloat[]{\includegraphics[width=7cm]{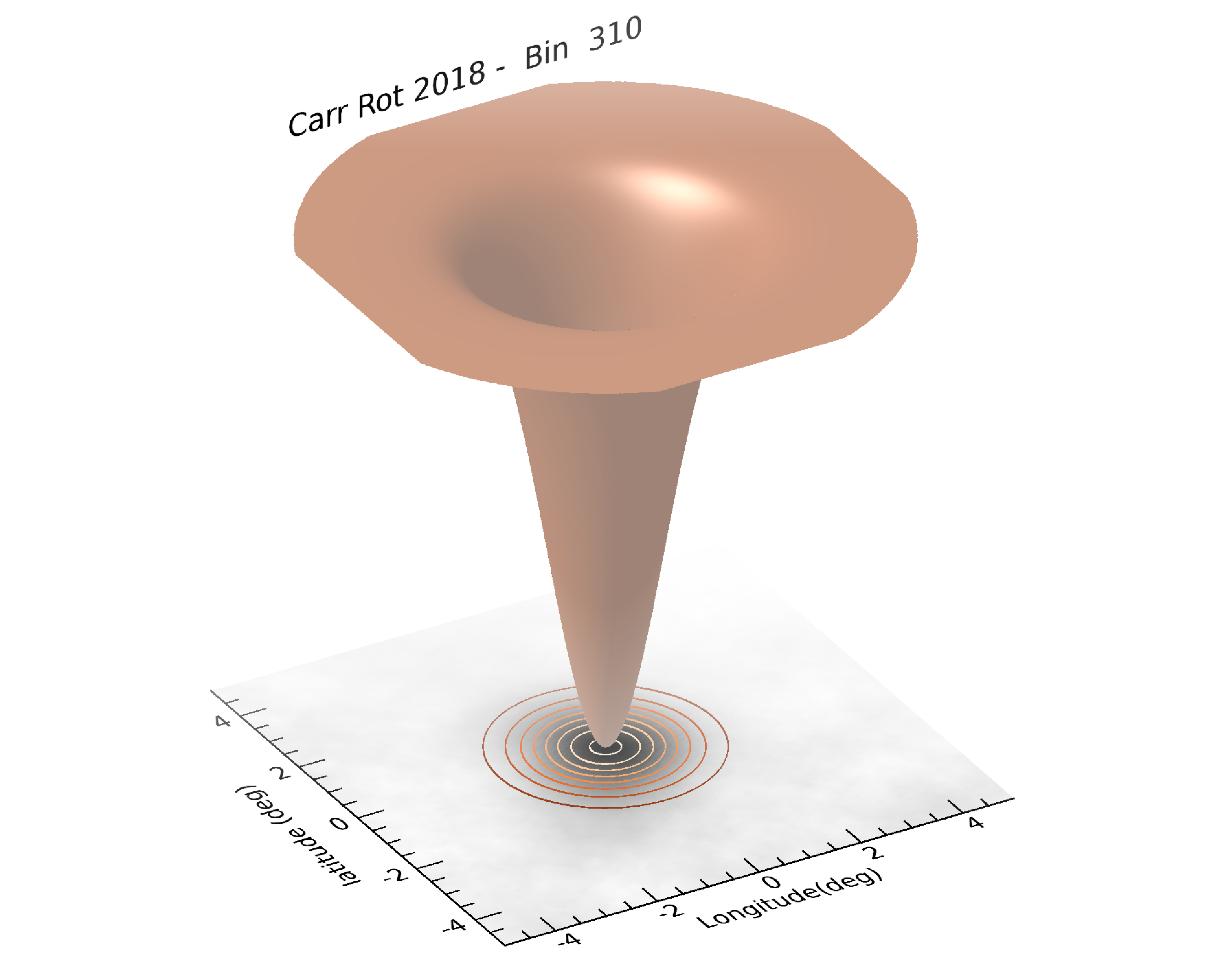}}%
    \caption{(a) Similar as Figure \ref{fig:maps16y17} but for the year 2018 and its fitted 2D gaussian (b).}%
    \label{fig:map18}%
\end{figure}

We have used standard HAWC procedures  \cite{2017ApJ...843...39A,2018ApJ...865...57A, 2005ApJ...622..759G} 
to construct sky maps in a region where the
Sun is positioned and the GCR deficit or Sun Shadow is caused by the interaction between these
GCR and both: the Sun itself and the magnetic field of the solar
atmosphere  \cite{2015ICRC...34...99E}. 
For this work, we
consider HAWC data taken during the years 2016 to 2018 and compute the  $SS_{RI}$  maps integrated
on a yearly basis, as shown in Figures
\ref{fig:maps16y17} (a and b), and  \ref{fig:map18} (a), 
 respectively.

\begin{figure}
\includegraphics[width=.4\textwidth]{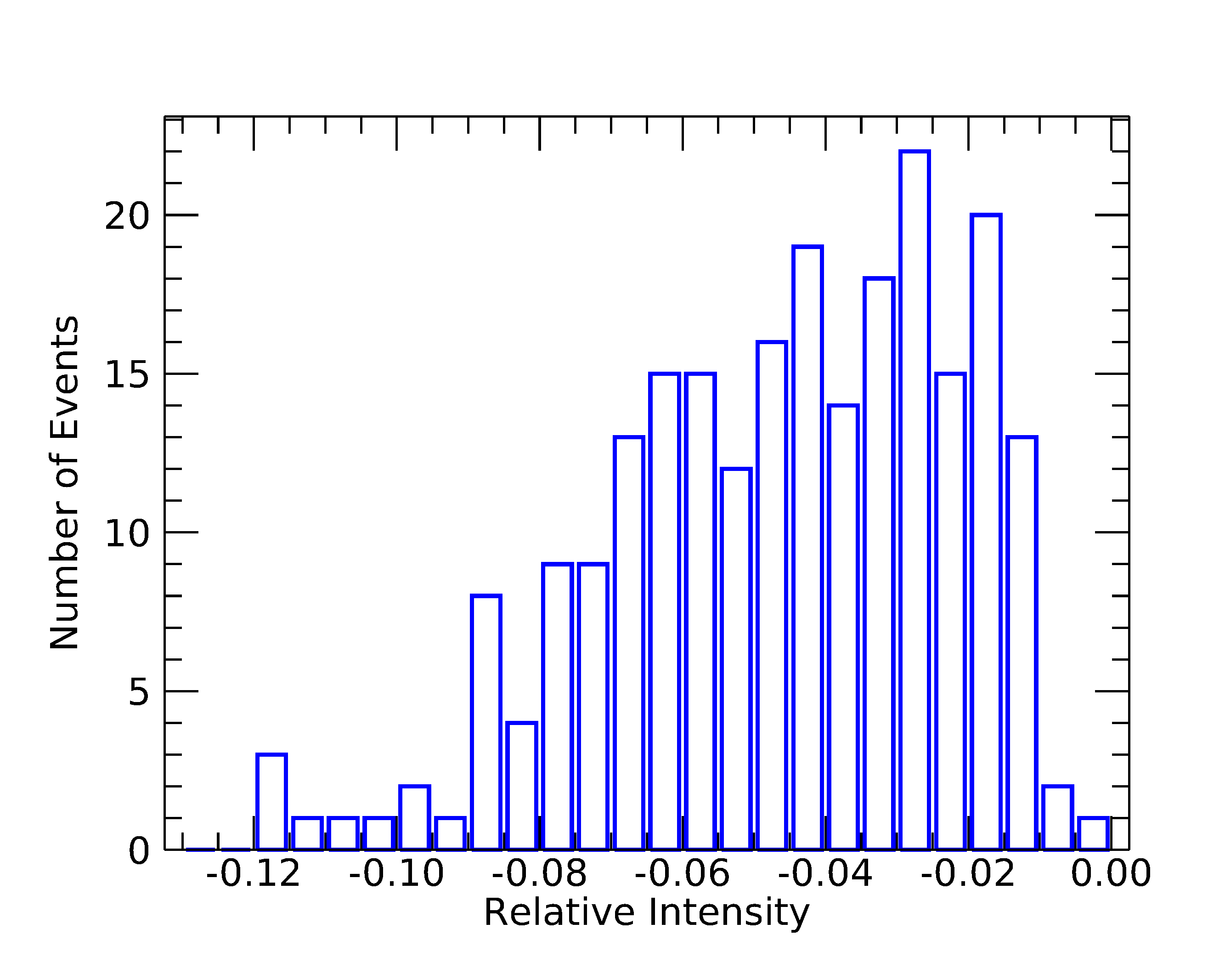}
     \caption{Distribution of the amplitudes of the  gaussians
       fitted to the GCR deficit maps.}
     \label{fig:hista}
   \end{figure}

\begin{figure}
  \centering
    \subfloat[2016]{{\includegraphics[width=7cm]{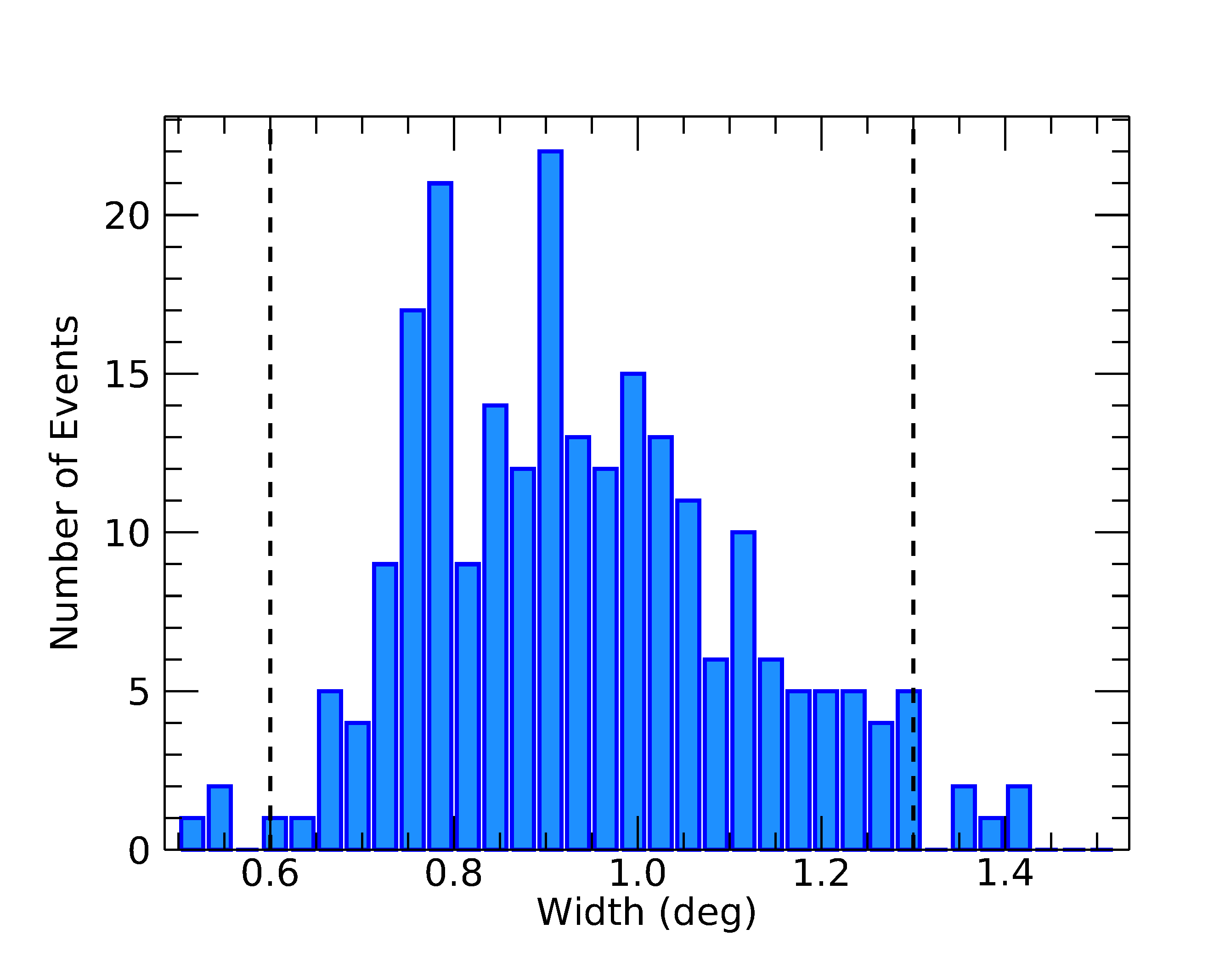} }}%
    \qquad
    \subfloat[2017]{{\includegraphics[width=7cm]{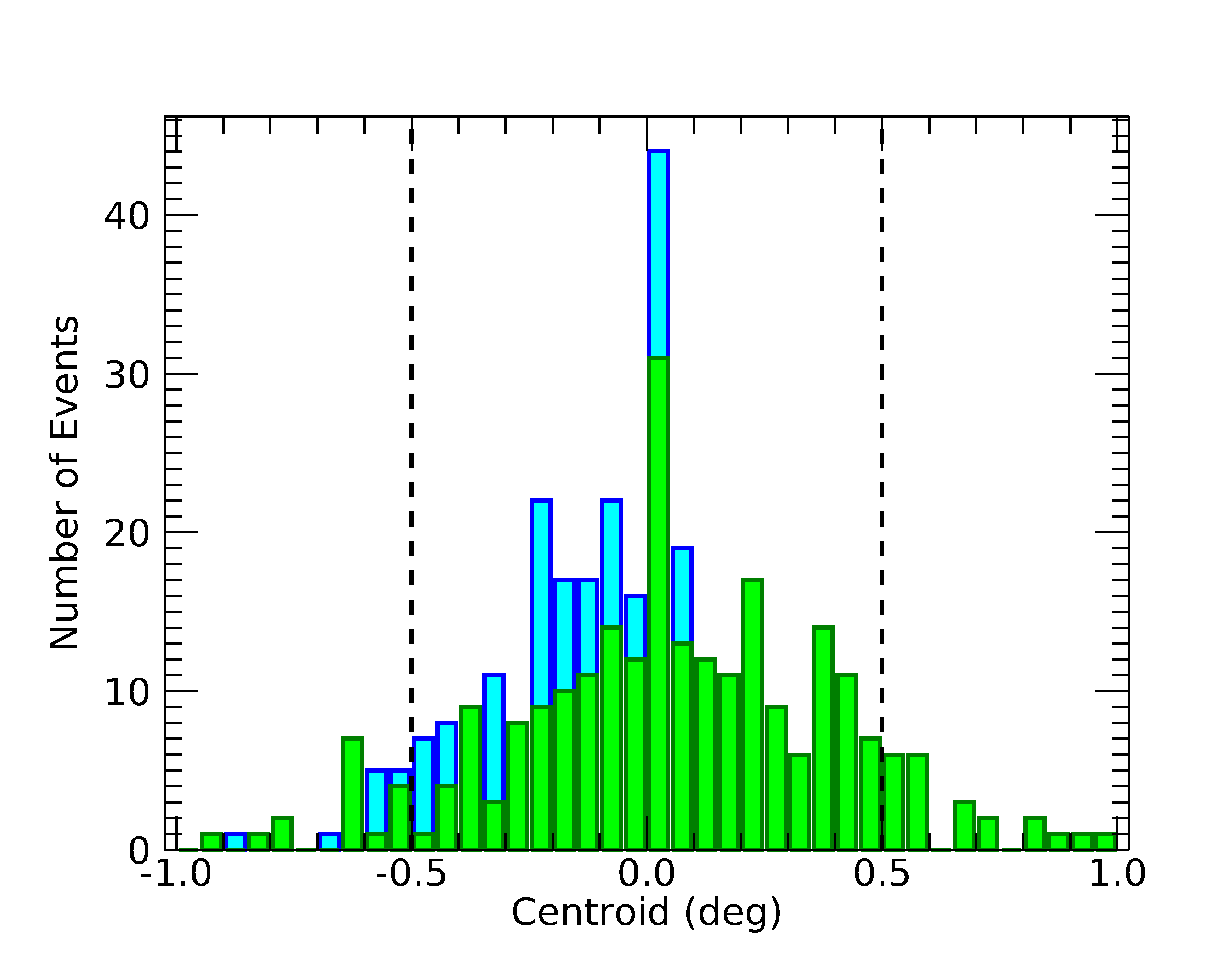} }}%
    \caption{Distribution of the widths (a) and centroid (b) of the  gaussians
       fitted to the GCR deficit maps.}%
    \label{fig:histwc}%
   \end{figure}

It is clear that the  $SS_{RI}$  changes with time. To order quantify these changes, we fitted a
circular 2-dimensional gaussian
$  F(x,y) = A_0 + A_{RI} \exp(- (   ((x-C_x)/W_x)^2 + ((y-C_y)/W_y)^2 )   /2) $
where  $W_x = W_y$. For example, Figure  \ref{fig:map18} (b) shows the 2D gaussian fitted to the  $SS_{RI}$   map integrated
for the year 2018. $A_{RI}$  corresponds to the relative intensity deficit measured by the height of the 2D
gaussian and this is the parameter selected for the analysis in this work. The $A_{RI}$ distribution is
plotted in Figure  \ref{fig:hista}. The distributions of the widths and centroids of
the fitted 2D gaussians are 
shown in Figure \ref{fig:histwc}, where the dashed vertical lines mark the cuts in both parameters that we
consider for further analysis. These are maps with
 $0.6 \le W \le 1.3$ and  $-0.5 \le
 C_{x,y} \le 0.5$.

As HAWC is located at $\sim 19^\circ$  North latitude, so the apparent latitudinal movement of the Sun
with respect to the observatory zenith angle  ($\theta_{z}$) changes during the year from  +4.5$^\circ$ to  -42.5$^\circ$, and
taking into account that the observational limit of HAWC is $\sim 40^\circ$, this implies that during six
months of the year (April to September) the $\theta_{z} $ is $< 20^\circ$.  We consider this to be the best time
window for our study. Therefore, we limit our study to Carrington rotations within this time
window. Furthermore, as the detected GCR flux decreases
as  $\sim\cos({\theta_{az}})$, for this work we assume that our uncertainty  ($ \epsilon$, marked by the error bars in the
following Figures) is:
$ \epsilon =\sqrt{\epsilon_{RI}^2 + A_{RI} \cos({\frac{9}{2}\theta_{az} +
    \pi})^2}$
where $A_0$ and $\epsilon_{fit}$  are the fitted amplitude
and its associated error and the factor $\frac{9}{2}$ 
 maximizes the error when $\theta_{z} = 20 ^\circ$.

\subsection{Time Evolution} \label{yearly}

\begin{figure}
  \centering
    \subfloat[]{{\includegraphics[width=7cm]{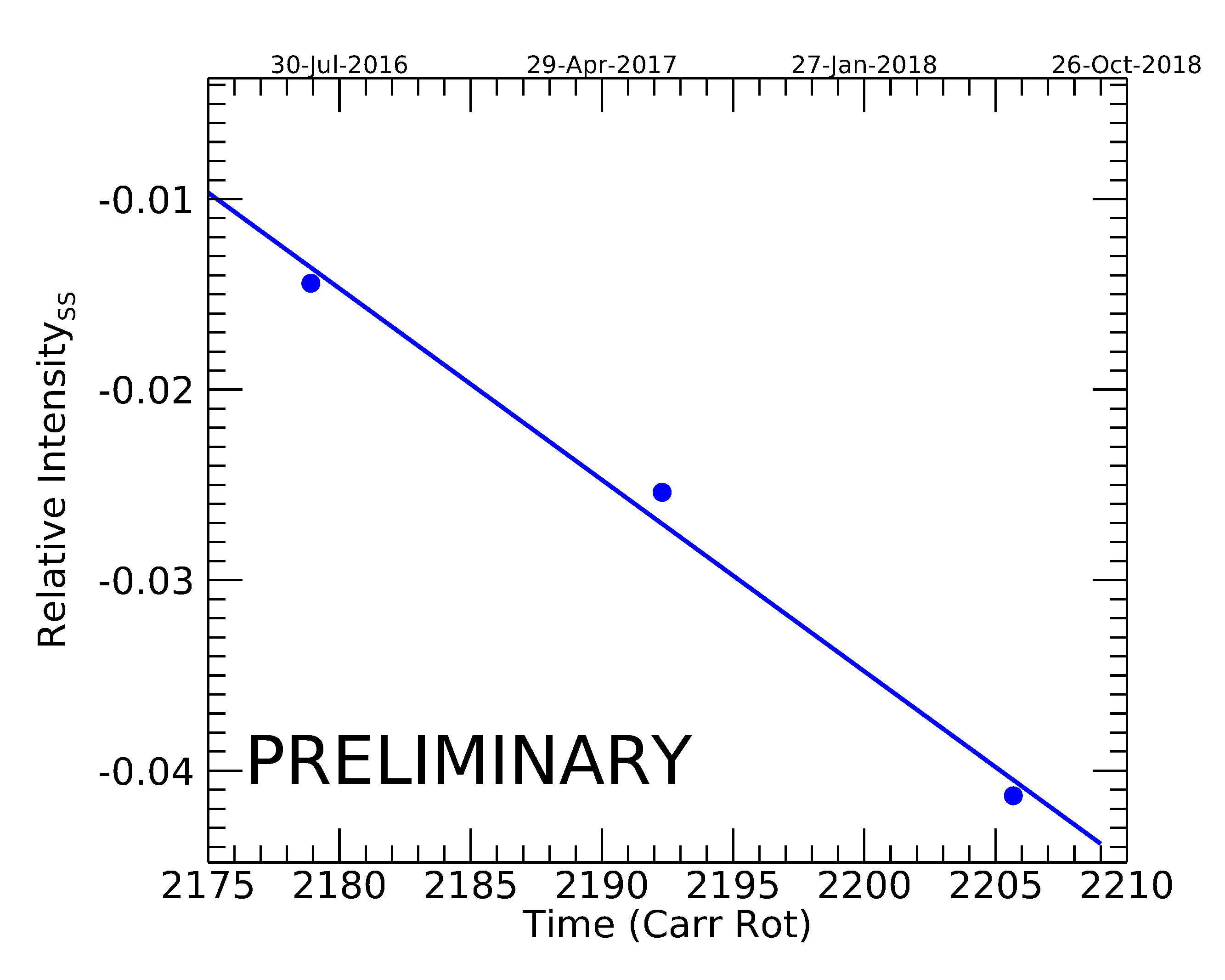} }}%
    \qquad
    \subfloat[]{{\includegraphics[width=7cm]{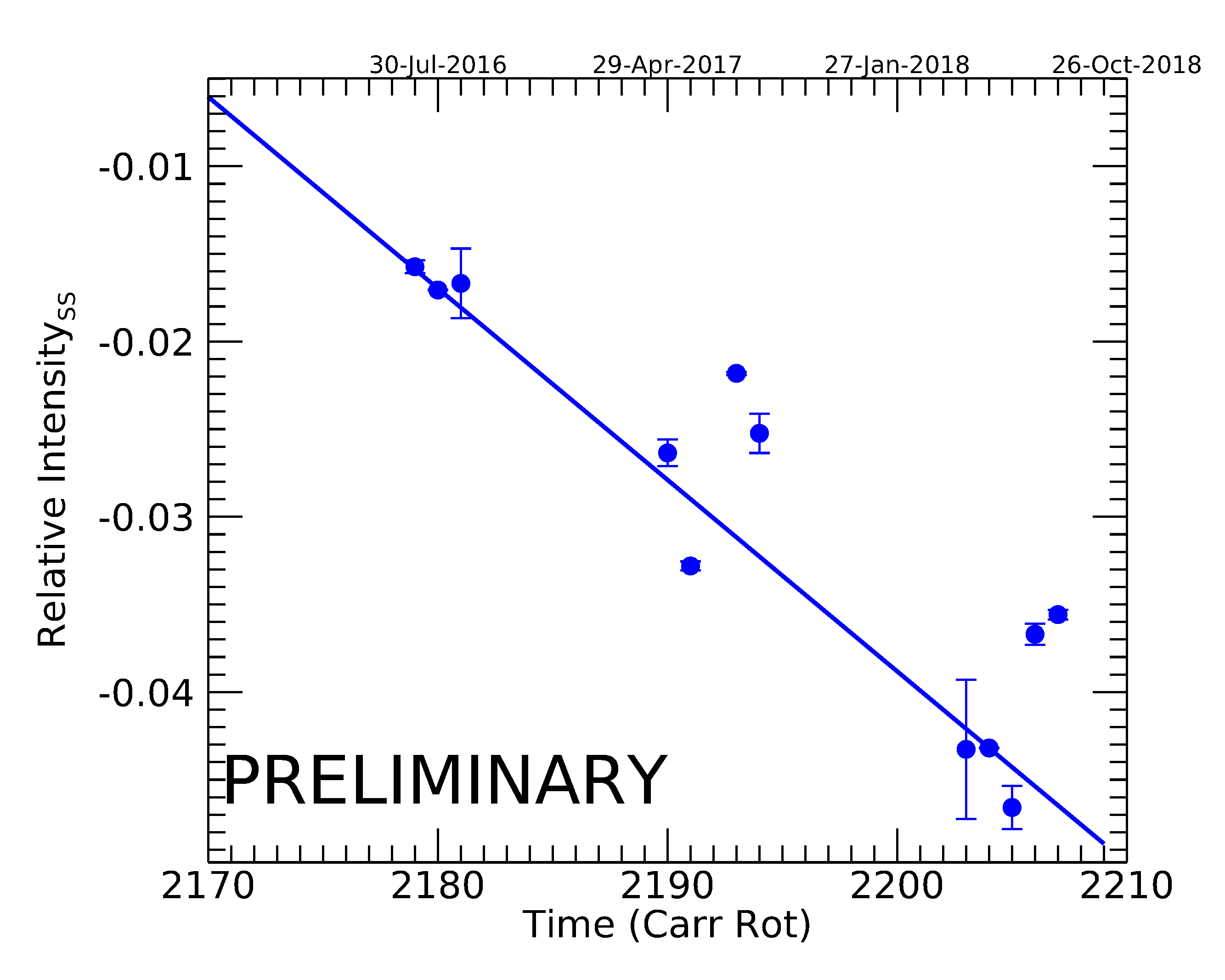} }}%
    \caption{Relative Intensity of the  $SS_{RI}$  as a function of time,
      integrated by six months (a) and one Carr\_Rot (b).}%
    \label{fig:ri_tim}%
   \end{figure}

The relative intensity decreases over the time during our period of
analysis as shown in Figure  \ref{fig:ri_tim} (a)
 where we show the variation with time of the relative intensity maps integrated over one
year. The high sensitivity of HAWC allows us to compute  $SS_{RI}$   maps with relatively short
integration times. In this work, we have generated maps with integration times of one solar
rotation, i. e., we have computed a map for each Carrington Rotation  (Carr\_Rot)  in our period of
study, starting with Carr\_Rot 2173 (Jan 21, 2016) up to Carr\_Rot 2210 (Oct 26, 2018). Clearly,
the time evolution of the $SS_{RI}$ integrated by Carr\_Rot also decreases along the time as shown in
Figure \ref{fig:ri_tim} (a) where the  $A_{RI}$ during each Carr\_Rot are plotted.

Further more the dependence of the RI is quasi-linear with the time with a rate of change of
-0.013 and -0.015 year$^{-1}$ for the yearly and Carr\_Rot integrated maps, respectively.

It is well know that the GCR flux is modulated by the solar activity. Although this modulation
has been observed at relatively low energies (up to several tens of GeV) and is attributed to the
solar wind and the large scale disturbances traveling in it. At higher energies ($\sim 10 - 200$ TeV), as
the HAWC maps analyzed in this work, the modulation should occur only very close to the solar
surface where the magnetic field is strong enough to deviate such high energy particles.

\section{Solar Cycle and Photospheric Magnetic Field} \label{sec:cycle}

\begin{figure}
  \centering
    \subfloat[]{{\includegraphics[width=7cm]{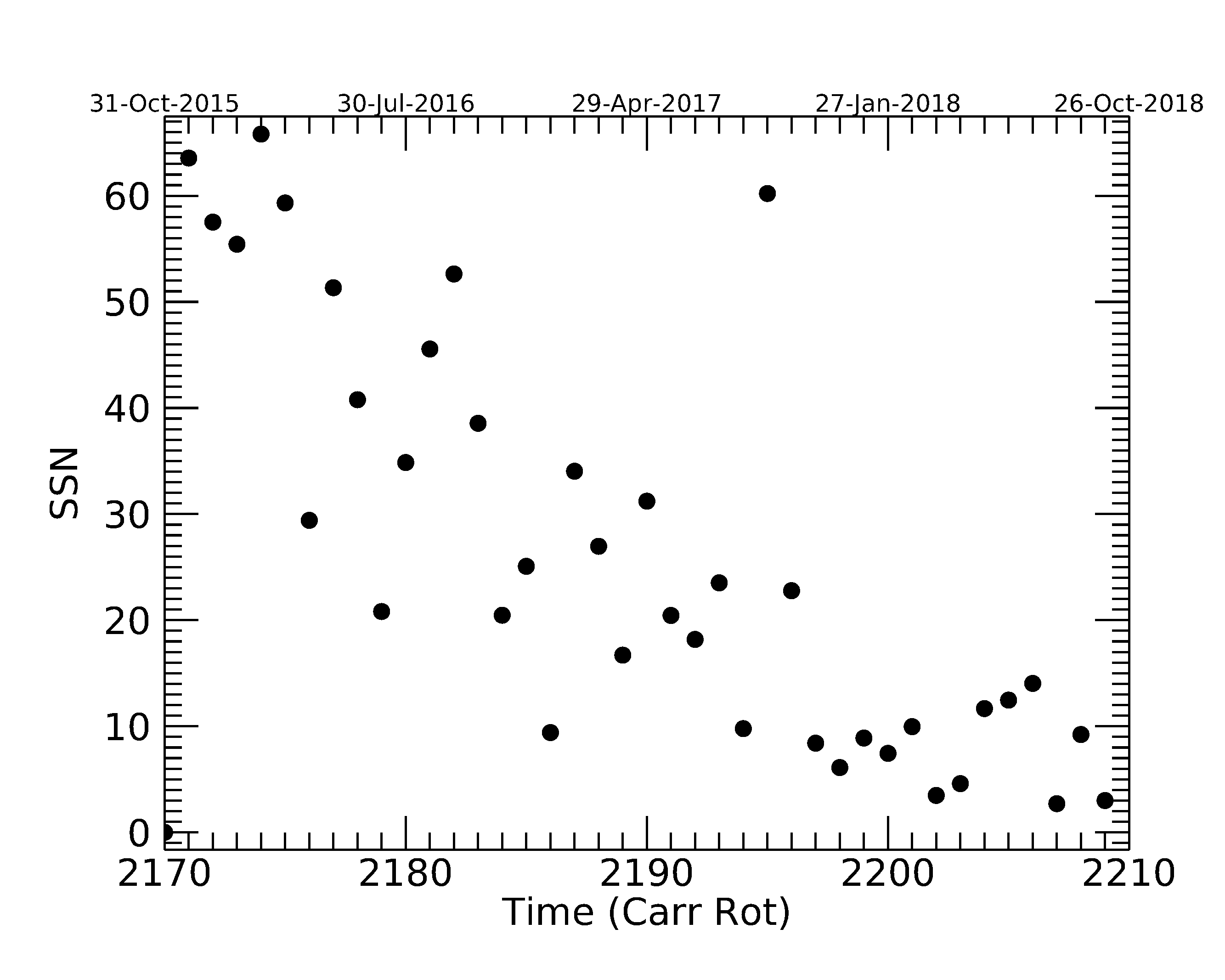} }}%
    \qquad
    \subfloat[]{{\includegraphics[width=7cm]{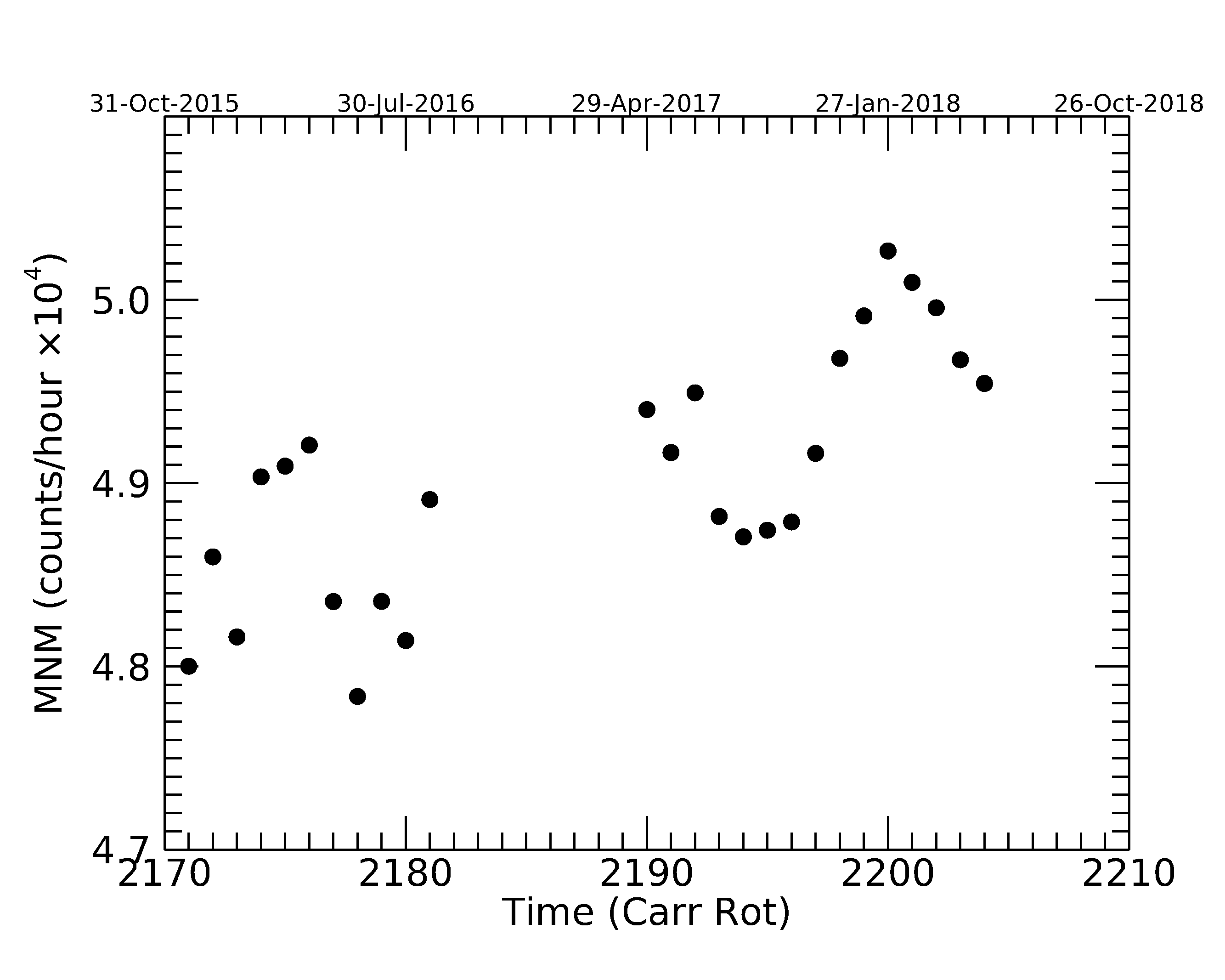} }}%
    \caption{SSN during solar cycle 24 (a), and the modulated GCR flux
      as observed by Neutron Monitors during the same period (b).}%
    \label{fig:ssn}%
   \end{figure}

The solar activity follows a magnetic cycle of  $\sim$  22 years, and during the 2016-2018 period,
the solar cycle 24 was declining as shown by the Sun Spot Number (SSN) plotted in Figure \ref{fig:ssn} (a).
In similar way, the modulated CGR flux measured by neutron monitors changes during the cycle
as shown in Figure \ref{fig:ssn} (b).

The SSN is a proxy of the solar activity, and should not be compared directly with the GCR
modulation, rather the magnetic field (and the solar wind speed for low energy GCR) should be
used. This represents a problem for low energy GCR studies, due to the fact that our
measurements of the solar wind parameters are limited to one or few points in the heliosphere,
where spacecraft reside.
Conversely, the high energy GCRs that HAWC observes are related to strong magnetic fields in the
low solar atmosphere, which are measurable (in the case of the photosphere) and can be
extrapolated up to few solar radii with a magnetic potential (current
free) model.

\begin{figure}
  \centering
    \subfloat[]{{\includegraphics[width=7cm]{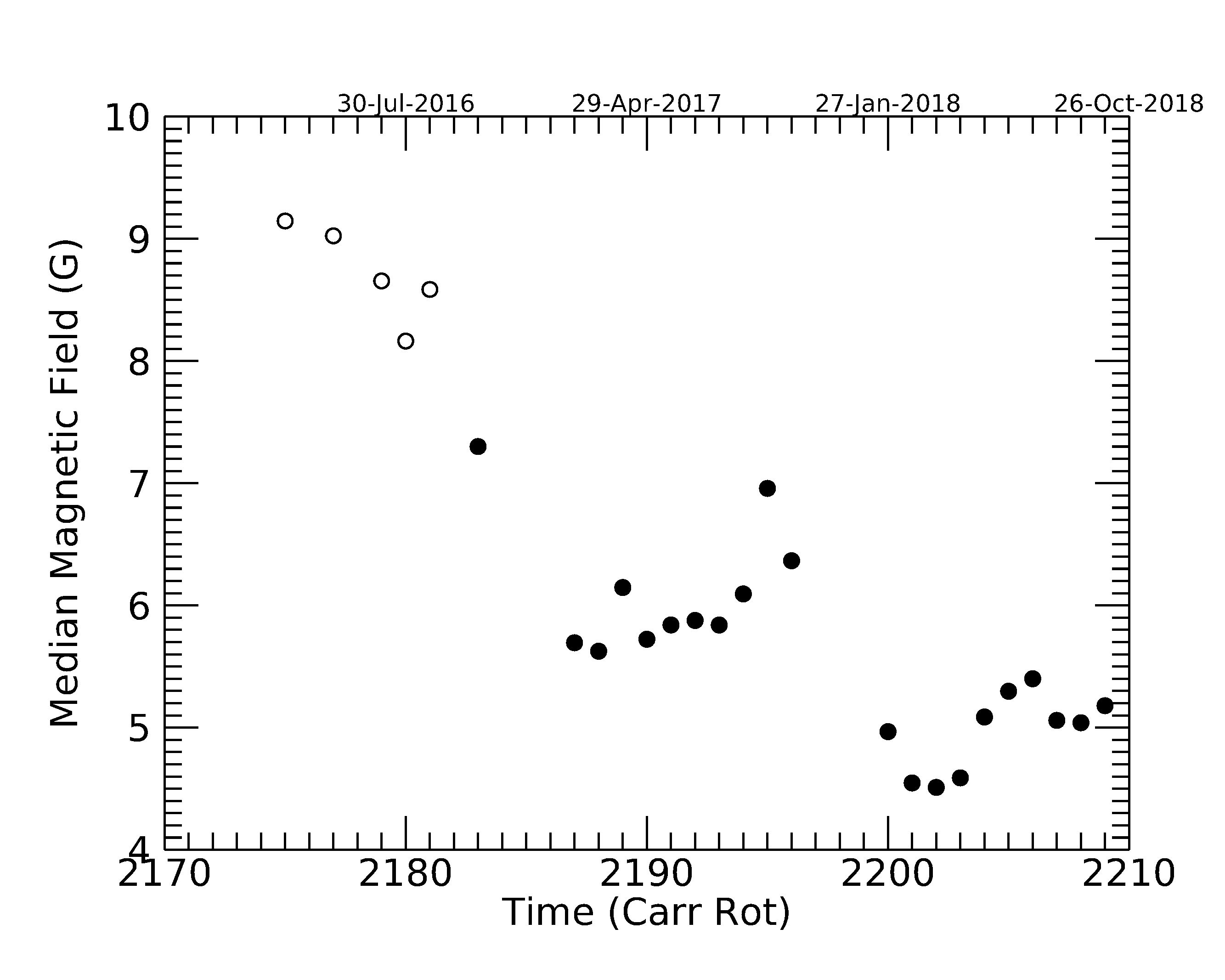} }}%
    \qquad
    \subfloat[]{{\includegraphics[width=7cm]{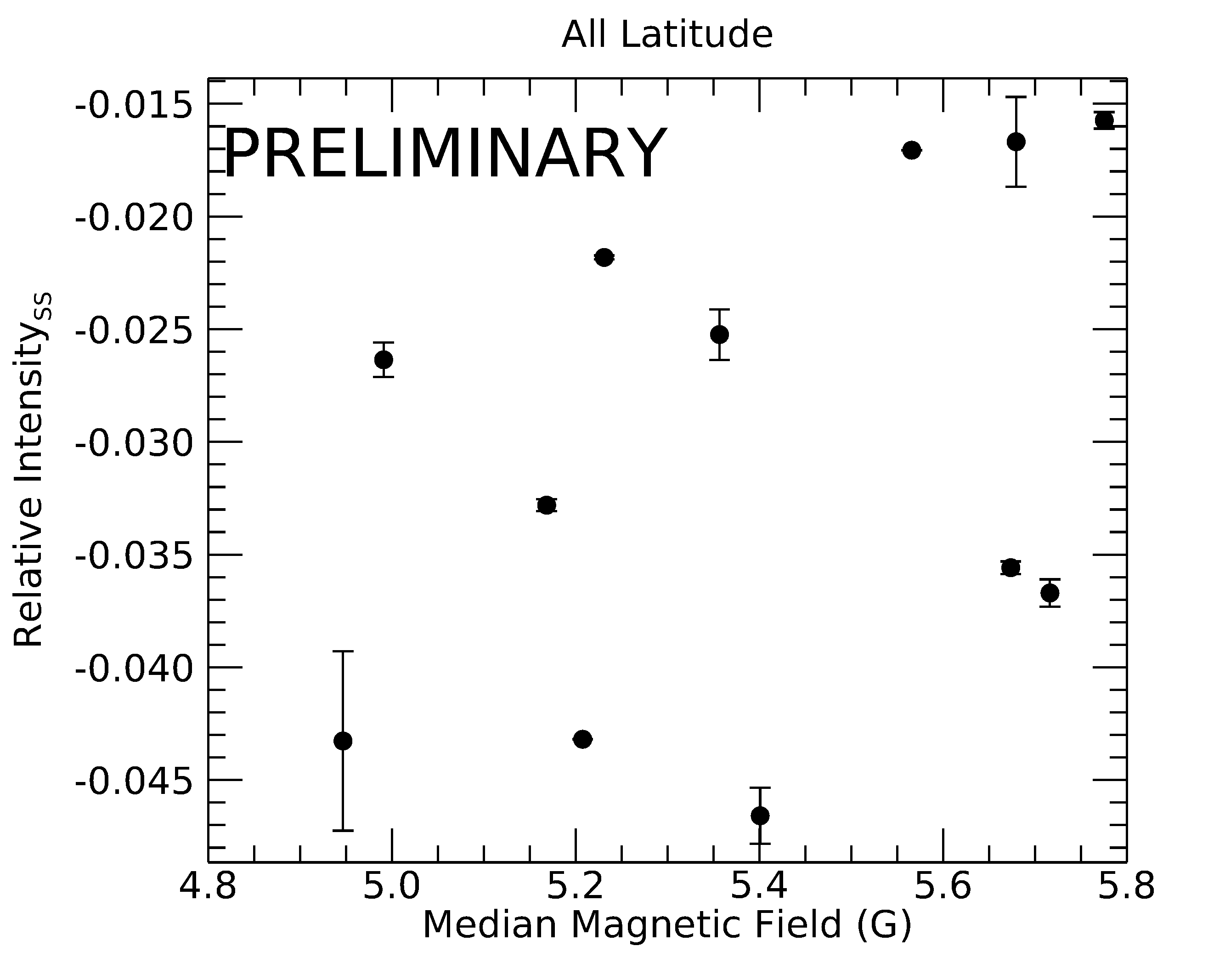} }}%
    \caption{(a) Median photospheric magnetic field computed at
      all latitudes for each  Carr\_Rot considered in this work. (b)
      The $A_{RI}$ as a function of the median photospheric magnetic field.}%
    \label{fig:magt}%
   \end{figure}

In this work we compare the changes in the $SS_{RI}$  with the photospheric magnetic field. Because
this field is highly variable, from $\sim 1000$  G inside Active
Regions to  $\sim 1$  1 G in quiet regions, we
use the median to characterize this field. Figure   \ref{fig:magt} (a)  shows the evolution of the median
photospheric magnetic field as a function of time during our period of study. To compare this
evolution with that of the   $SS_{RI}$, Figure  \ref{fig:magt} (b)  shows the RI of the $SS_{RI}$ as a function of the median
photospheric magnetic field.

We attribute this lack of correlation to the fact that the solar magnetic field evolves differently,
depending on the latitude. For example, in the so called Active Region belt (ranging from
 $\sim -30^\circ$ to $\sim
30^\circ$ 
of latitude), the magnetic field attains the highest strengths during the maximum of activity
with a toroidal configuration, wheras this strength is lowest during the minimum of activity. On the
other hand the configuration on the poles (lat $\ge \pm 60^\circ$) attains the the lowest values during the
maximum and increases towards the minimum of activity with a global dipolar topology.

\begin{figure}
  \centering
    \subfloat[]{{\includegraphics[width=7cm]{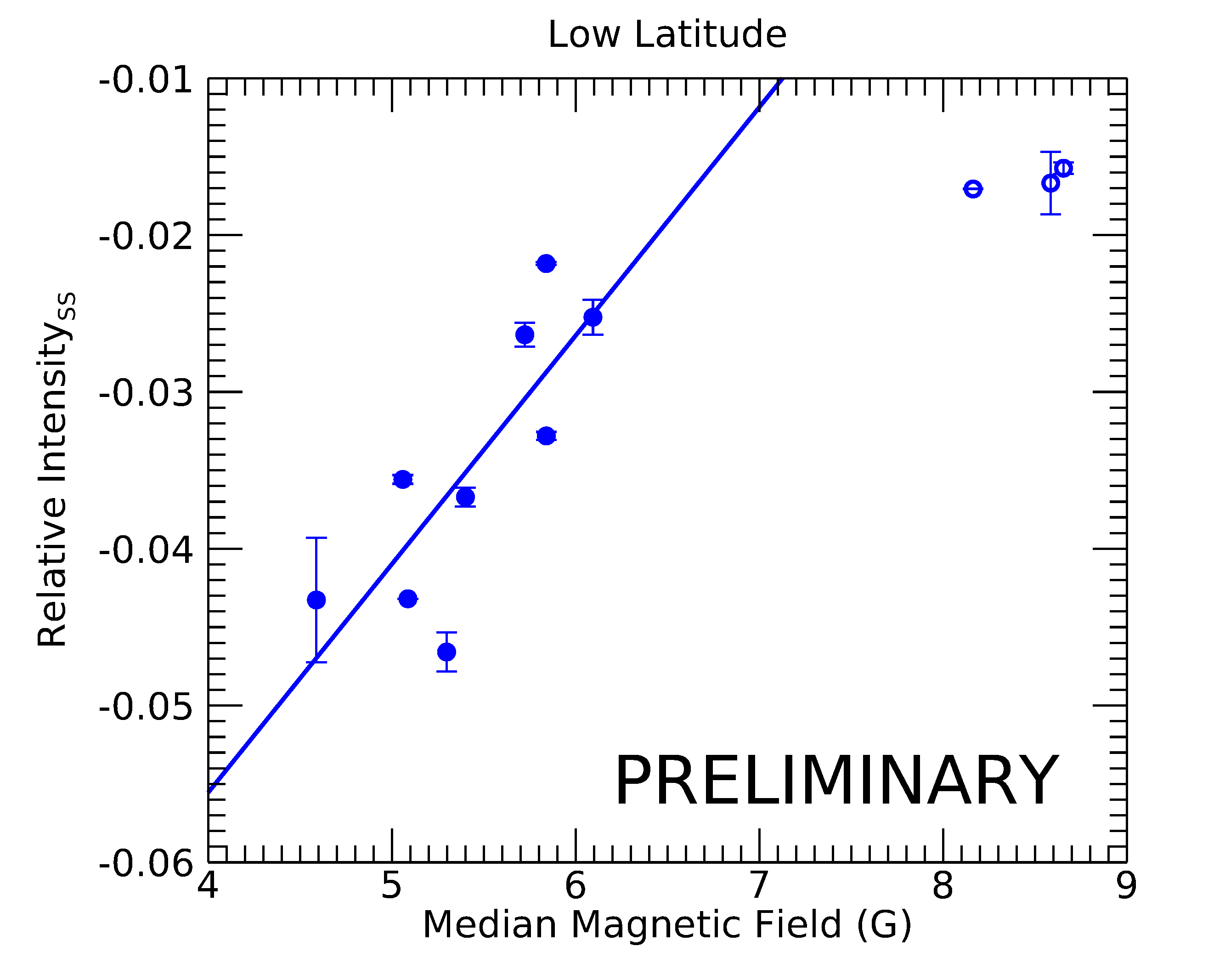} }}%
    \qquad
    \subfloat[]{{\includegraphics[width=7cm]{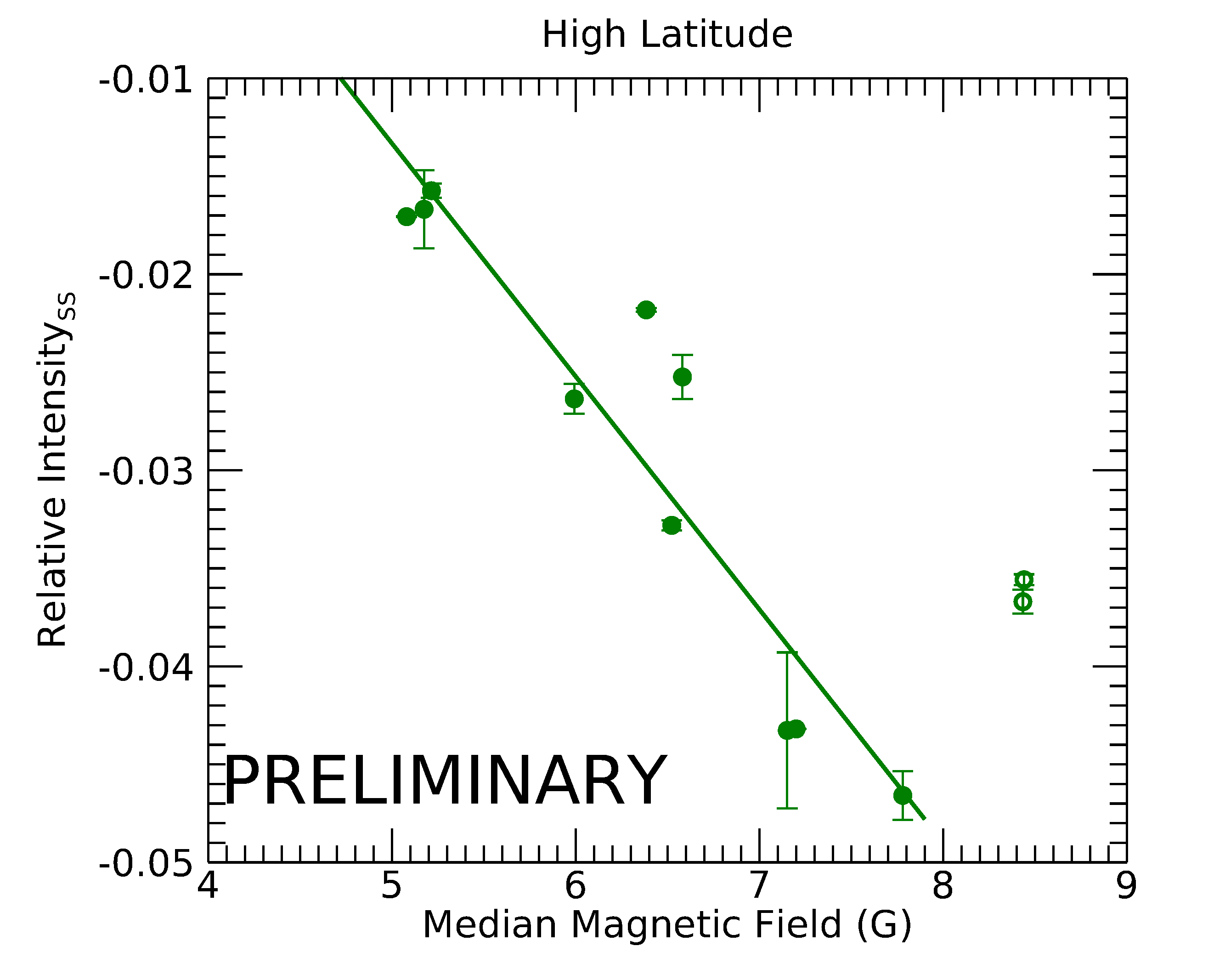} }}%
    \caption{The $A_{RI}$ as a function of the median photospheric magnetic
      field at low  (a) and high (b) latitudes.}%
    \label{fig:maglh}%
   \end{figure}

The RI as a function of the toroidal (low latitudes) and poloidal (high latitudes) magnetic fields
are shown in Figures  \ref{fig:maglh} a and b, respectively.
Interestingly, the dependence (at low magnetic field values), of the RI with the magnetic field
is linear at both low and high latitudes, but of opposite sign. The
corresponding rates of change are
$1.46 \times 10^{-2}$ G$^{-1}$
and 
$-1.19 \times 10^{-2}$ G$^{-1}$.

This relationship is valid for low activity periods, where the median magnetic field is lower
than 8 G. When the solar activity and the magnetic field is greater, as during Carr\_Rots 2179 , 2180
and 2181 (on July and August 2016) the linear relationship is no
longer valid, as seen in Figure \ref{fig:maglh}.
 
\section{Bending of cosmic rays in coronal magnetic field }  \label{sec:sim}

A simulation was performed to estimate the GCR bending in the solar corona. In this approximation,
the solar magnetic field was modeled assuming a symmetric photospheric field
($B_{phot}$) with a range of 0.1 to 20 G and decreasing as the square
of the radial distance
 ($B_{r} =
 \frac{B_{phot}}{(r-r_{phot})^2}$).
 The Larmor radius of a CR in the magnetic field is
given as,
 \begin{equation}
R_L ~~ = ~~( 3.3 \times 10^{12} ~ cm ) \times\frac {E (GeV) }{ B( \mu G) }
\end{equation}

\begin{figure}
  \centering
    \subfloat[]{{\includegraphics[width=7cm]{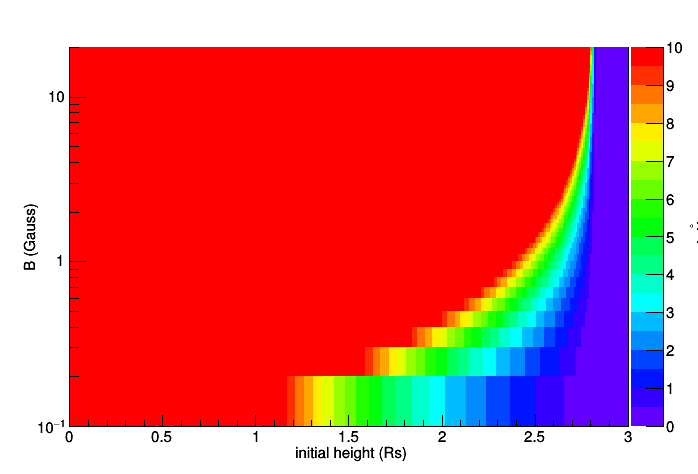} }}%
    \qquad
    \subfloat[]{{\includegraphics[width=7cm]{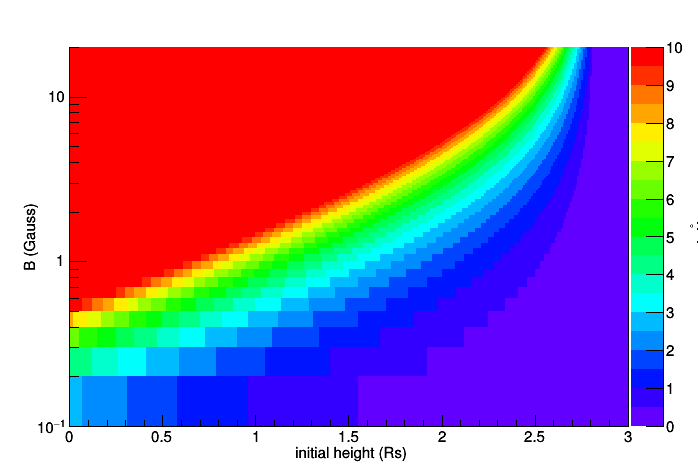} }}%
    \caption{Simulations of the deviation of GCR of 7 (left) and 85
      (right) TeV, passing through the
      coronal magnetic field. }%
    \label{fig:sim}%
   \end{figure}

The deviation experienced by the cosmic ray while it is traveling a
small distance '$dL$'  through
the magnetic field can be expressed in terms of the Larmor radius as $
d\Theta = dL/R_L$.  The total
deflection angle of the particle was estimated by a numerical integration over its trajectory within
the magnetic field profile as described above. The simulation was carried out for different energies;
tracing the trajectory of the particle inside the corona; defining the initial distance as the distance
between the photosphere and the line marked by the initial trajectory of the particle along a radius
perpendicular to this line; and setting the initial and final integration points the positions where the
particle crosses the 3 $R_{\odot}$ limit. As an example, Figure
\ref{fig:sim}  shows the deviation angle (indicated by the
color code) as a function of the initial distance and the photospheric magnetic field for two
energy bins of HAWC, 7 and 85 TeV on the left and right panels respectively. It is clear that the deviation angle is
less than 2$^\circ$ when particles of 7 and 85 TeV traverse the atmosphere at an initial distance of 2 and
0.3 solar radii, respectively.

\section{Summary and Conclusions} \label{sec:concl}

We performed a temporal analysis of the deficit of GCR caused by the Sun during the 2016
to 2018 year period. In particular, we analyzed the behavior of the amplitude of a 2D symmetric
gaussian, adjusted to the relative intensity maps of the  $SS$  made using two integration times of $\sim 1$ 
and 6 months. We found a decreasing trend of the $SS_{RI}$ with
time. Furthermore, by
comparing the $SS_{RI}$  with the photospheric magnetic field, we found a linear correlation between the
signal amplitude and the median magnetic field measured at the Active Region belt  ($-30^\circ \ge$ Lat $\ge
30^\circ$),
i.e., with the toroidal magnetic field during the high activity phase of the cycle, and a linear
anti-correlation between the $SS_{RI}$ and the median magnetic field at polar regions  ( Lat $
\ge \pm 60^\circ$). 
This is related to the transition from a multipolar configuration to a simpler dipolar magnetic field
during the low activity phase. Finally, we presented simulations of the passage of high energy
particles through a spherical symmetrical magnetic field, to show that the GCR of energies
detected by HAWC are deviated few degrees and cause the observed $SS$.

\section*{Acknowledgements}
A. Lara thnaks PASPA-UNAM for partial support.

\end{document}